\documentclass[prd,twocolumn,superscriptaddress,linenumbers]{revtex4-1}
\usepackage{graphicx}

\usepackage{hyperref}
\hypersetup{
 colorlinks
}

\def\aap{Astron.\ Astrophys.\ }
\def\mnras{Mon.\ Not.\ R.\ Astron.\ Soc.\ }
\def\apjl{Astrophys.\ J.\ Lett.\ }

\begin{document}

\title{New constraints on primordial black holes abundance from
  femtolensing of~gamma-ray bursts}

\author{A.~Barnacka}
\email[e-mail:~]{abarnack@camk.edu.pl}
\affiliation{Nicolaus Copernicus Astronomical Center, Warszawa, Poland}
\affiliation{DSM/IRFU/SPP, CEA/Saclay, F-91191 Gif-sur-Yvette, France}
\author{J.-F.~Glicenstein}
\email[e-mail:~]{ glicens@cea.fr}
\affiliation{DSM/IRFU/SPP, CEA/Saclay, F-91191 Gif-sur-Yvette, France}
\author{R.~Moderski}
\affiliation{Nicolaus Copernicus Astronomical Center, Warszawa, Poland}

\date{\today}

\begin{abstract}
The abundance of primordial black holes is currently significantly
constrained in a wide range of masses.  The weakest limits are
established for the small mass objects, where the small intensity of
the associated physical phenomenon provides a challenge for current
experiments.  We used gamma-ray bursts with known redshifts detected by the
Fermi Gamma-ray Burst Monitor (GBM) to search for the femtolensing
effects caused by compact objects.  The lack of femtolensing detection
in the GBM data provides new evidence that primordial black holes in
the mass range $5 \times 10^{17}$ -- $10^{20}\,$g do not constitute
a major fraction of dark matter.  
\end{abstract}

\maketitle


\section{Introduction}
%
%
Dark matter is one of the main and most challenging open problems in
cosmology or particle physics, and a number of candidates for
particle dark matter have been proposed over the years~\cite{feng10}.
An alternative idea that the missing matter consists of compact
astrophysical objects was first proposed in the
1970s~\cite{1974MNRAS.168..399C,1974Natur.248...30H,1971MNRAS.152...75H}.
An example of such compact objects are primordial black holes (PBHs)
created in the very early Universe from matter density perturbations.

Recent advancements in experimental astrophysics, especially the
launch of the FERMI satellite with its unprecedented sensitivity
has revived the interest in PBH
physics~\cite{2010PhRvD..81j4019C,2011PhRvL.107w1101G}.  One of the
most promising way to search for PBHs is to look for lensing
effects caused by these compact objects.  
Since the Schwarzschild radius of PBH is comparable to the photon wavelength, 
the wave nature of electromagnetic radiation has to be taken into account.  
In such a case, lensing caused by PBHs introduces an interferometry pattern in
the energy spectrum of the lensed object~\cite{1981SvAL....7..213M}.
The effect is called 'femtolensing'~\cite{1992ApJ...386L...5G} due to
very small angular distance between the lensed images.  The phenomenon
has been a matter of extensive studies in the
past~\cite{2001ASPC..239....3G},
but the research was
almost entirely theoretical since no case of femtolensing has been
detected as yet.  Gould~\cite{1992ApJ...386L...5G} first suggested
that the femtolensing of gamma-ray bursts (GRBs) at cosmological distances
could be used to search for dark matter objects in the mass range
$10^{17} - 10^{20}\,$g.
Femtolensing could also be a signature of another
dark matter candidate: clustered axions \cite{1996ApJ...460L..25K}.

In this paper, we present the results of a femtolensing search performed on 
the spectra of GRBs with known redshifts detected by the Gamma-ray Burst Monitor
(GBM) on board the FERMI satellite.  
The non observation of femtolensing on these bursts provides new constraints
on the PBHs fraction in the mass range $5 \times 10^{17} -
10^{20}\,$g.  We describe in details the optical depth
derivation based on simulations applied to each burst
individually.  The sensitivity of the GBM to the femtolensing
detection is also calculated.

The paper is organized as follows: in Sec.~\ref{sec:estimate} the basic equations 
for femtolensing and the calculation of lensing probability are given.  Section~\ref{sec:data}
describes the data sample and simulations.  In
Sec.~\ref{sec:results} the results are presented, while
Sec.~\ref{sec:conclusions} is devoted to discussion and
conclusions.


\section{Femtolensing\label{sec:estimate}}


\subsection{Magnification and spectral pattern}
%
Consider a lensing event of a GRB by a compact object.  The angular diameter
distances from the observer to the lens, from the lens to the GRB
source and from the observer to the source are $D_{OL}$, $D_{LS}$, and
$D_{OS}$, respectively.  Coordinates are taken in the lens plane.
The lens, with mass $M$, is located at the origin.  The source position
projected onto the lens plane is given by $r_{S}$, the distance
between the lens and true source position.  The Einstein radius
$r_{E}$ is given by
\begin{eqnarray}
r_E^2 &=& \frac{4 G M}{c^2} \frac{D_{OL}D_{LS}}{D_{OL}+D_{LS}}
\nonumber \\ &\approx& {\left( c \times 0.1 \,\mbox{s} \right)}^2
\left( \frac{4\xi - 4}{\xi^2} \right) \left(
\frac{D_{OS}}{5\,\mbox{Gpc}} \right) \left( \frac{M}{10^{19}
  \,\mbox{g}} \right)\,,
 \label{rE}
\end{eqnarray}
where
\begin{equation}
\xi = \frac{D_{OS}}{D_{OL}} \,.
 \label{D}
\end{equation}
The image positions are given as usual by 
\begin{equation}
r_{\pm}=  \frac{1}{2}(r_S \pm \sqrt{r_S^2 + 4 r_E^2}) \,.
 \label{rplus}
\end{equation}
The time delay $\delta t$ between the two images is given by
\begin{equation}
c \delta t = V(r_{+};r_S) - V(r_{-};r_S) \,,
 \label{dt2}
\end{equation}
where $V(r;r_{S})$ is the Fermat potential at the position
$r$ in the lens plane. One finds:
\begin{equation}
c \delta t= \frac{1}{2} \left( \frac{1}{D_{LS}}+ \frac{1}{D_{OL}}
\right) (r_{+}^2 - r_{-}^2) - \frac{4GM}{c^2}
\ln(\frac{r_{+}}{r_{-}}) \,.
 \label{dt3}
\end{equation}
The phase shift between the two images is 
$\Delta \phi~=~E~\delta t/\hbar$, 
where $E$ is the energy of the photon.

In the case of a point source, the amplitude contributed by the
$r_{\pm}$ images is
\begin{equation}
A_{\pm} \propto \frac{\exp(i\phi_{\pm}) }{\sqrt{ |
    1-\frac{r_E^4}{r_{\pm}^4} |}} \,.
\label{A}
\end{equation}

\noindent
The magnification $A^{2}$ is obtained by summing the
amplitudes~(\ref{A}) and squaring, which gives
\begin{eqnarray}
|A|^2 &=& |A_{+}+A_-|^2 = \nonumber \\
&=& \frac{1}{ 1-\frac{r_E^4}{r_{+}^4} }
+\frac{1}{1-\frac{r_E^4}{r_{-}^4}} + \frac{2\cos(\Delta
  \phi) }{\sqrt{| 1-\frac{r_E^4}{r_{+}^4} |} \sqrt{|
    1-\frac{r_E^4}{r_{-}^4} |}} \,.
 \label{Mag}
\end{eqnarray}

The energy dependent magnification produces fringes in the energy spectrum of the lensed object.   

In principle, the finite size of the source and the relative motion of
the observer, the lens and the source have to be taken into account.
If the GRB is observed at a time $t_{expl}$ after the beginning of the
burst, its size projected onto the lens plane is
\begin{eqnarray}
s_{GRB} &\approx& \frac{D_{OL}}{D_{OS}} \frac{c t_{expl}}{\Gamma}
\nonumber \\ &\approx& \left( c \times 0.005\,\mbox{s} \right) \, \left(
\frac{t_{expl}}{1\,\mbox{s}} \right) \left( \frac{\Gamma}{100} \right)^{-1}
\left( \frac{\xi}{2} \right)^{-1} \,,
\label{sgrb}  
\end{eqnarray}
where $\Gamma$ is the Lorentz factor of the burst. Note that the
Lorentz factor of GRBs is estimated to be in excess of $100$, so that
$s_{GRB}$ given in Eq.~(\ref{sgrb}) is overestimated.

The ratio of $s_{GRB}$ to $r_{E}$ is therefore
\begin{eqnarray}
\frac{s_{GRB}}{r_{E}} & \approx & 0.05\, (\xi-1)^{-\frac{1}{2}} \left(
 \frac{t_{expl}}{1 \,\mbox{s}} \right) \left( \frac{\Gamma}{100}
 \right)^{-1} \nonumber \\ && \times \left( \frac{D_{OS}}{5
 \,\mbox{Gpc}}\right)^{-\frac{1}{2}} \left( \frac{M}{10^{19}
  \,\mbox{g}}\right)^{-\frac{1}{2}} \,.
\label{finitegrb}
\end{eqnarray}
Equation (\ref{finitegrb}) shows that the finite size of the GRB can
be in general safely neglected if $t_{expl} < 10\,\mbox{s}$.

The Einstein radius crossing time $t_{E}$ is given by
\begin{eqnarray}
t_{E} &=& \frac{r_E}{v} \nonumber \\ &\approx& 100 \,\mbox{s} \,
\left( \frac{r_{E}}{c \times 0.1 \,\mbox{s}} \right)
  \left(\frac{v}{300\,\mbox{km/s}} \right)^{-1} \,,
\label{einsteincrossingtime}
\end{eqnarray}
where $v$ is the projected velocity of the source in the lens plane.
Equation (\ref{einsteincrossingtime}) shows that $t_{E} \gg t_{expl}$
under reasonable assumptions on the velocities.  If so, the motion of
the source in the lens plane can also be neglected.  In the analysis
of GRB spectra, it is thus assumed that the point source -- point lens
assumption is valid and that the source stays at a fixed position in
the lens plane.

\subsection{Lensing probability\label{sec:probability}}
%

The lensing probability of gamma ray burst events is calculated in two steps. 
First, the optical depth $\tau$ for lensing by compact objects is calculated 
according to the formalism of Fukugita et al. \cite{1992ApJ...393....3F}. 
The cosmological parameters used in the calculation are:
a mean mass density $\Omega_{M}=0.3$ and a normalized cosmological constant $\Omega_{\Lambda}=0.7.$ 
The calculations are made for both the Friedmann-Lemaître-Robertson-Walker (FLRW) 
and the Dyer-Roeder \cite{1973ApJ...180L..31D} cosmology. 
In our sample, the GRB redshift $z_{s}$ is known. The lens redshift $z_{L}$ is 
assumed to be given by the maximum of the $d\tau/dz_{L} (z_{S})$ distribution
(see e.g. Fig.~5 of \cite{1992ApJ...393....3F}). 
When $\tau \ll 1,$ the lensing probability $p$ is given by 
$p = \tau \sigma$ where $\sigma$ is the ``lensing cross-section'' (see Chap. 11 of \cite{1992grle.book.....S}). 
    
In this paper, the cross-section is defined in the following way. Fringes are 
searched in the spectra of GRBs. These fringes are detectable only for certain
positions $r_{S}$ of the source. The exact criteria for detectability will be 
given in section \ref{sec:simulations}. The maximum and minimum position  of $r_{S},$ in units 
of $r_{E}$ are noted $r_{S,min}$ and $r_{S,max}.$ They are found by simulation 
and depend on the GRB redshift and luminosity.
 A minimum value of $r_S$ occurs because the period of the spectral fringes becomes larger than the GBM energy range at small $r_{S}$. 

The femtolensing ``cross-section'' is simply 
\begin{equation}
\sigma = r^{2}_{S,max}-r^{2}_{S,min}
 \label{sigma}
\end{equation}

The lensing probability does not depend on the individual masses of lenses, but only on the density of compact objects $\Omega_{CO}$. 
In the optical depth calculation, an increase in the mass of the lenses is compensated by a decrease in the number of scatterers.
Therefore, the constraints for a given mass depend only on the cross section $\sigma$. 

\section{Data Analysis\label{sec:data}} 

In our analysis, we use a sample of GRBs with known redshifts. 
The selection of these bursts is described in Sec.~\ref{sec:selection}.
Then each burst is fitted to a standard  spectral model, as explained in Sec.~\ref{sec:analysis}.
Finally, the sensitivity of each burst to femtolensing is studied with simulated data.
The simulation is described in Sec.~\ref{sec:simulations}.

\subsection{Data selection\label{sec:selection}}
The Gamma-Ray Burst detector (GBM)\cite{2009ApJ...702..791M} on-board the Fermi satellite 
consists of 12 NaI and 2 BGO scintillators 
which cover the energy range from 8 keV up to 40 MeV in 128 energy bins. 
These detectors perform a whole sky monitoring. In the first two years of operation, GBM triggered on roughly 500 GRBs.
In this paper, only the bursts with known redshifts 
have been investigated. 
The initial  sample consisted of 32 bursts taken from 
Gruber et al.~\cite{2011A&A...531A..20G} 
and 5 additional bursts from the GRB Coordinates Network (GCN) circulars\footnote{http://gcn.gsfc.nasa.gov}. 
 
For 17 burst the amount  of available data was not sufficient to obtain good quality spectra. 
The final sample thus  consists of 20 bursts, which are listed in the Tab.~\ref{table}.
 
\subsection{Data processing and spectral analysis\label{sec:analysis}}
 
The GBM data are publicly available in the CSPEC format and were downloaded from the Fermi FSSC website \footnote{http://fermi.gsfc.nasa.gov/ssc/}.
The CSPEC files contain the counts in 128 energy channels binned in 1.024 s for all detectors. 
Only detectors with a minimal signal to noise ratio of 5.5 in each bin were selected for the analysis.   

Data were analyzed with the {\tt RMfit}  version 33pr7 program.   
The {\tt RMfit} software package was originally developed for the time-resolved 
analysis of BATSE GRB data but has been adapted to GBM and other instruments.

For each detector with sufficient data, the background was subtracted and the 
counts spectrum of the first ten seconds of the burst (or less if the burst was shorter) was extracted. 

The energy spectrum was obtained with a standard forward-folding algorithm.
Several  GRB spectral models such as  a broken power law (BKN),  Band's model (BAND) or a smoothly broken power law (SBKN) where considered. 
The femtolensing effect was added as a separate model. 
The magnification and the oscillating fridges where calculated according to Eq.~(\ref{Mag}), 
then multiplied with the BKN or BAND functions.

\subsection{Simulations\label{sec:simulations}}

 The detectability of spectral fringes has been studied with 
simulated signals. The detectability depends on 
one side on the luminosity and the redshift of the bursts, 
and on the other side on the detector energy resolution and the data quality. 
The sensitivity of the GBM to the lens mass $M$ depends strongly also on the energy range and
resolution of the GBM detectors. When small masses are considered, 
the pattern of spectral fringes appears outside of the energy range. The large masses produce
fringes with hardly detectable amplitudes and periods smaller than the 
energy bin size. 

Because the data quality and the background are not easily simulated, 
the detectability estimation is performed on real data.
Namely, GRB events with known redshift are selected. Since the source redshift is known,
the lens redshift is assumed to be the maximum value of $d\tau/dz_{L} (z_{S})$ 
as explained in Sec. \ref{sec:probability}. For a given observed GRB, 
the femtolensing signal depends thus
only on 2 parameters: the lens mass $M$ and the source position in the lens plane $r_{S}.$ 
The data are then processed  as follows: 
\begin{enumerate}
\item The magnification (Eq.~\ref{Mag}) as a function of the energy is calculated for the given
lens mass $M$ and position of the source $r_{S}.$ 
\item This magnification is then convolved  with the instrumental resolution matrix to obtain magnification factors for each channel of the detector.  
\item The spectral signal is extracted from the data by subtracting the background. It is then
multiplied by the corrected magnification. 
\item The background is added back.
\end{enumerate}

The detectability calculation can be illustrated with the luminous  burst GRB090424592. 
The spectral data of this burst were first fitted with standard spectral models: BKN, 
SBKN and BAND.  
The GRB090424952 burst is best fitted with the BAND model. The fit has $\chi^{2} = 78$ for 67 degrees of freedom (d.o.f). 
The BAND model has 4 free parameters: the amplitude A, the low energy spectral index $\alpha$, the high energy spectral index $\beta$ and the peak energy $E_{peak}$ \cite{2012arXiv1201.2981G}.

\begin{figure}
 \includegraphics[width=6cm,angle=-90]{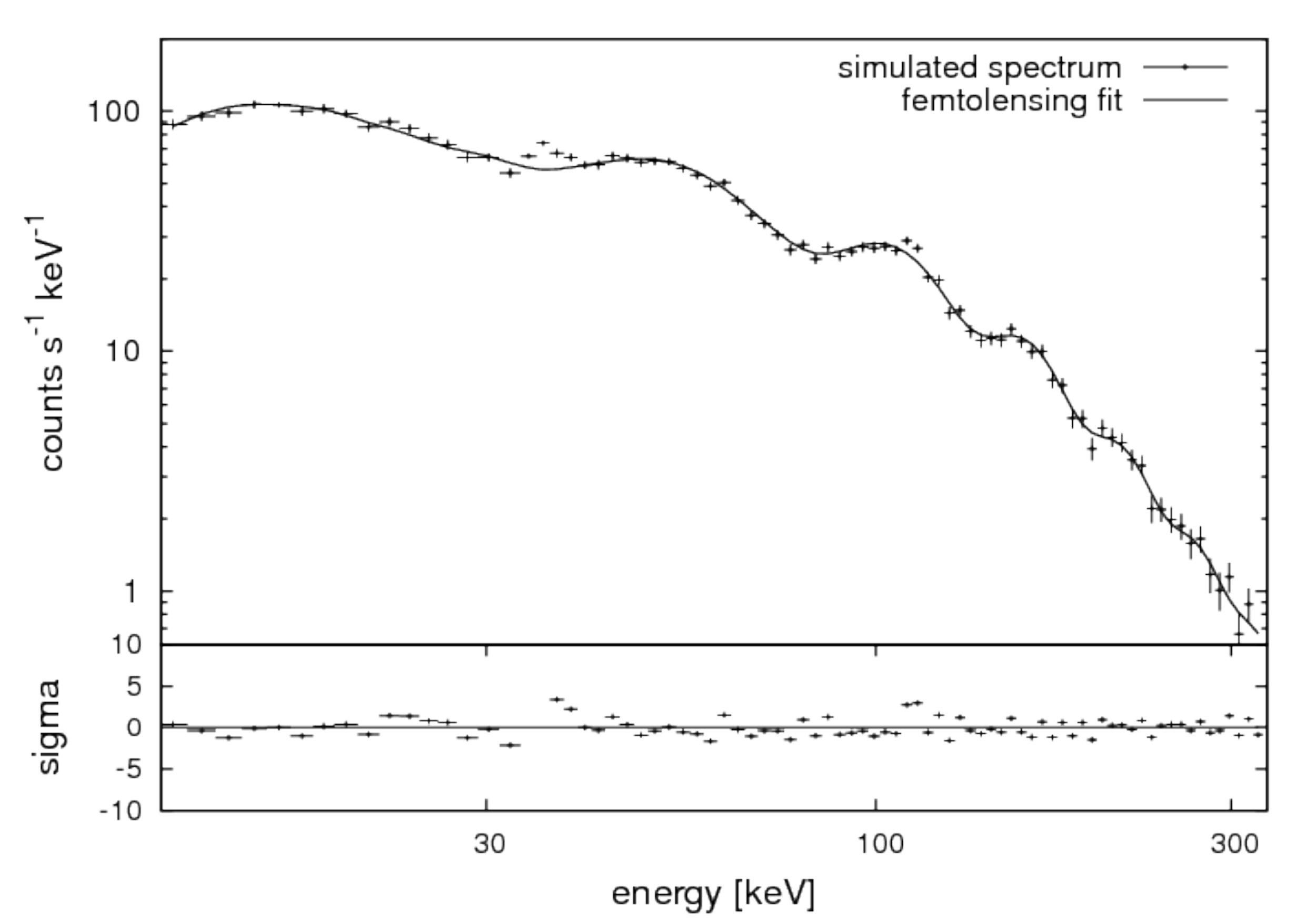}%
 \caption{\label{MD_femto} Simulated spectrum obtained with GRB 090424592.  
The spectrum  was fitted with femtolensing$+$BAND model.
The fit has $\chi^2$ = 92 for 73 d.o.f. 
The fit parameters are: $A=0.34\pm0.02\,$ph$\,$s$^{-1}\,$cm$^{-2}\,$keV$^{-1}$, $E_{peak}=173\pm12\,$keV, 
$\alpha=-0.84\pm0.03$ and $\beta=-3.9\pm7.5$. 
The simulated femtolensing effect is caused by a lens at redshifts $z_L=0.256$ and a source at $z_S=0.544$.
The simulated mass is $M = 5 \times 10^{18} \,$ g and the mass reconstructed from the fit is $5.8 \times 10^{18}\,$g.
 The source is  simulated at position $r_S=2$. The position reconstructed from the fit is $r_S =1.9$.}  
 \end{figure}

\begin{figure}
 \includegraphics[width=6cm,angle=-90]{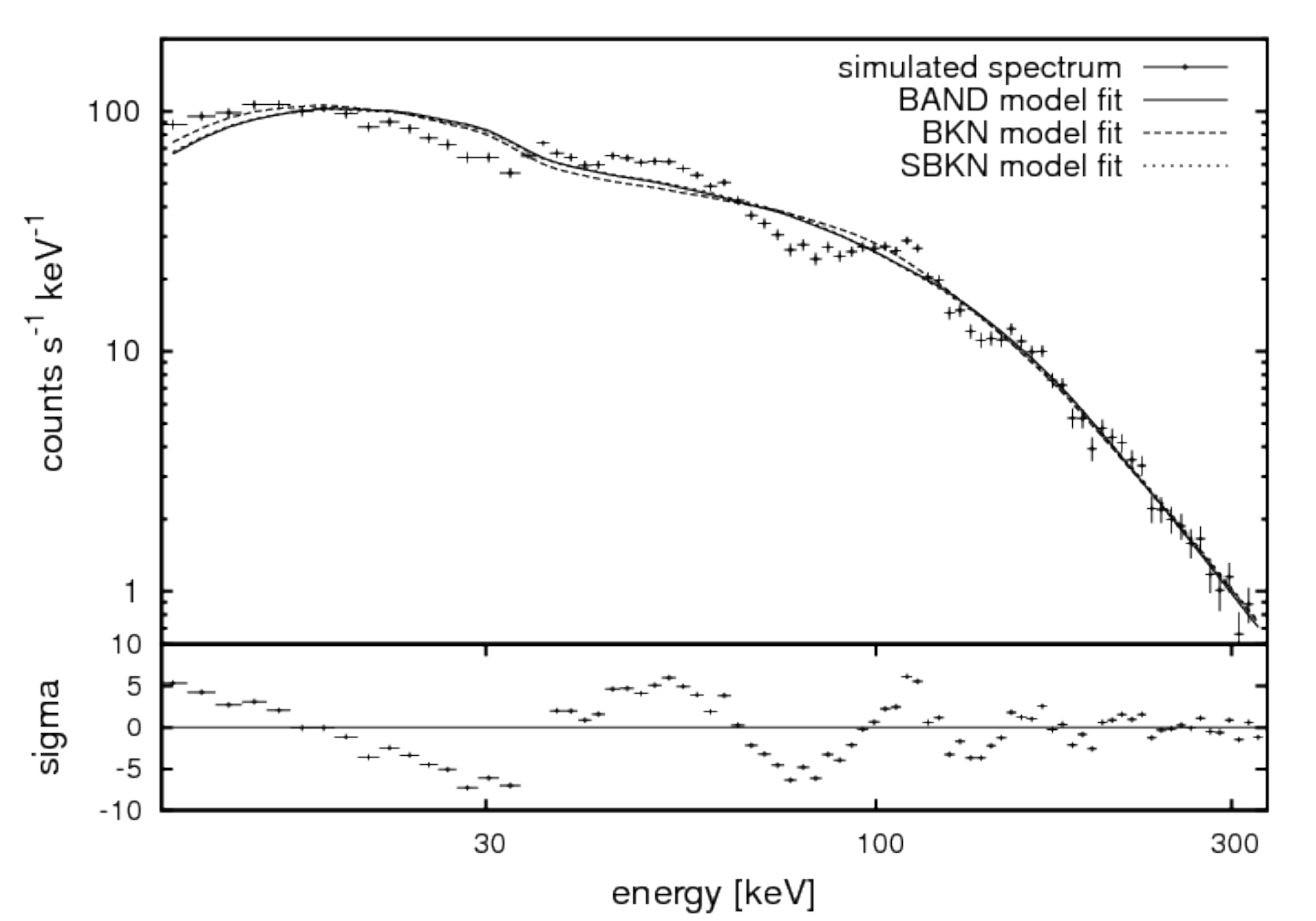}%
 \caption{\label{MD_BAND}  Simulated femtolensed spectrum fitted with the BAND model. 
 The fit has $\chi^2 = 810$ for $75$ d.o.f. 
 The fit parameters are: 
 $A= 0.5\pm0.03\,$ph$\,$s$^{-1}\,$cm$^{-2}\,$keV$^{-1}$,  $E_{peak}=147\pm5\,$keV,
 $\alpha=-0.58\pm0.03$ and $\beta=-2.4\pm0.1$.
 The SBKN model fit is almost indistinguishable from the BKN model fit.}
 \end{figure}

\begin{figure}
   \includegraphics[width=6cm,angle=-90]{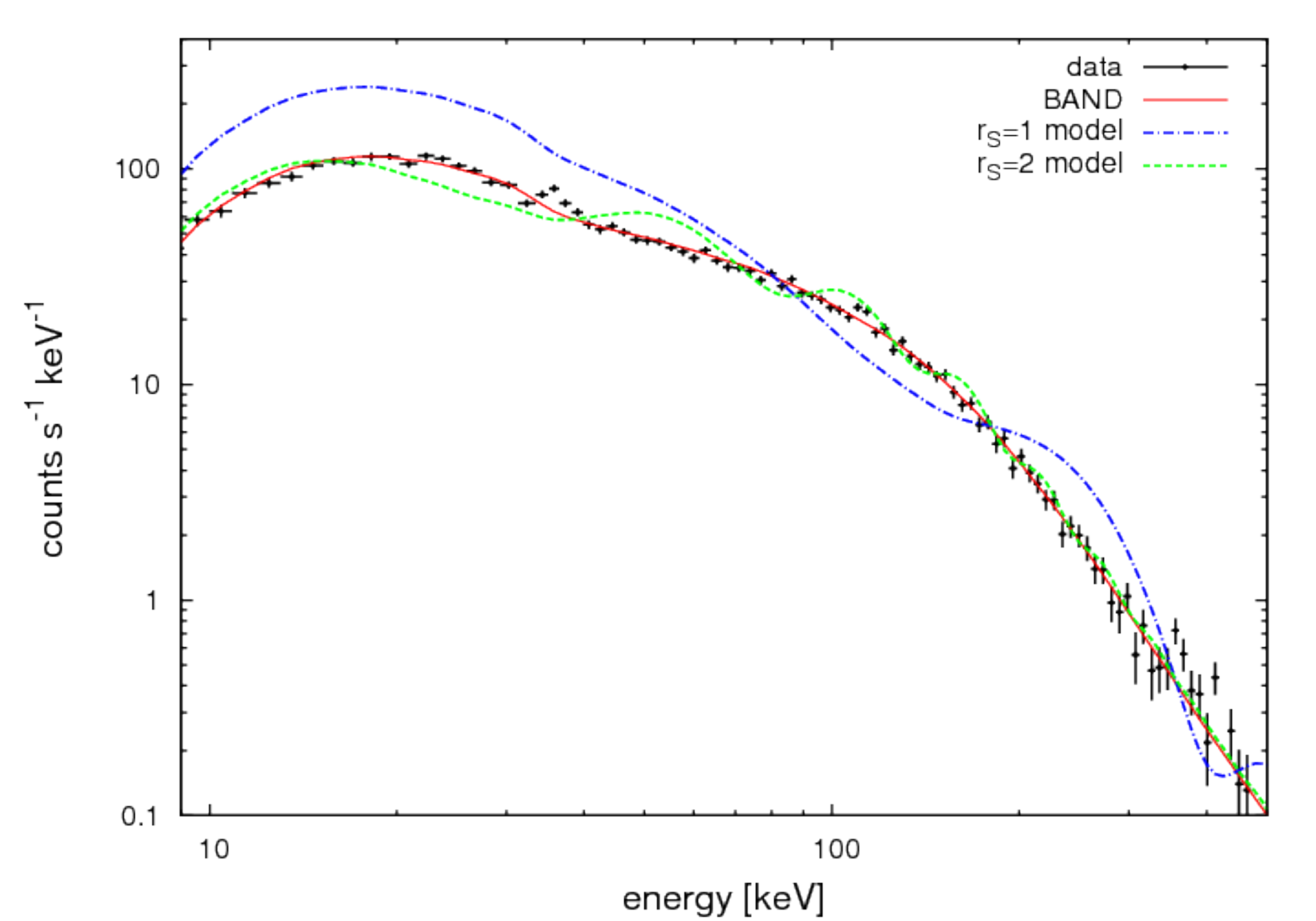}%
   \caption{\label{rS12} The spectrum of GRB 090424592 using NaI
     detector n7, with the BAND and femtolensing fits superimposed.
      The parameters are $r_{S} = 1$, $2$, and lens mass $5\times10^{18} \,$g.
     The models are convolved with the response matrix. }
 \end{figure}

\begin{figure}
   \includegraphics[width=6cm,angle=-90]{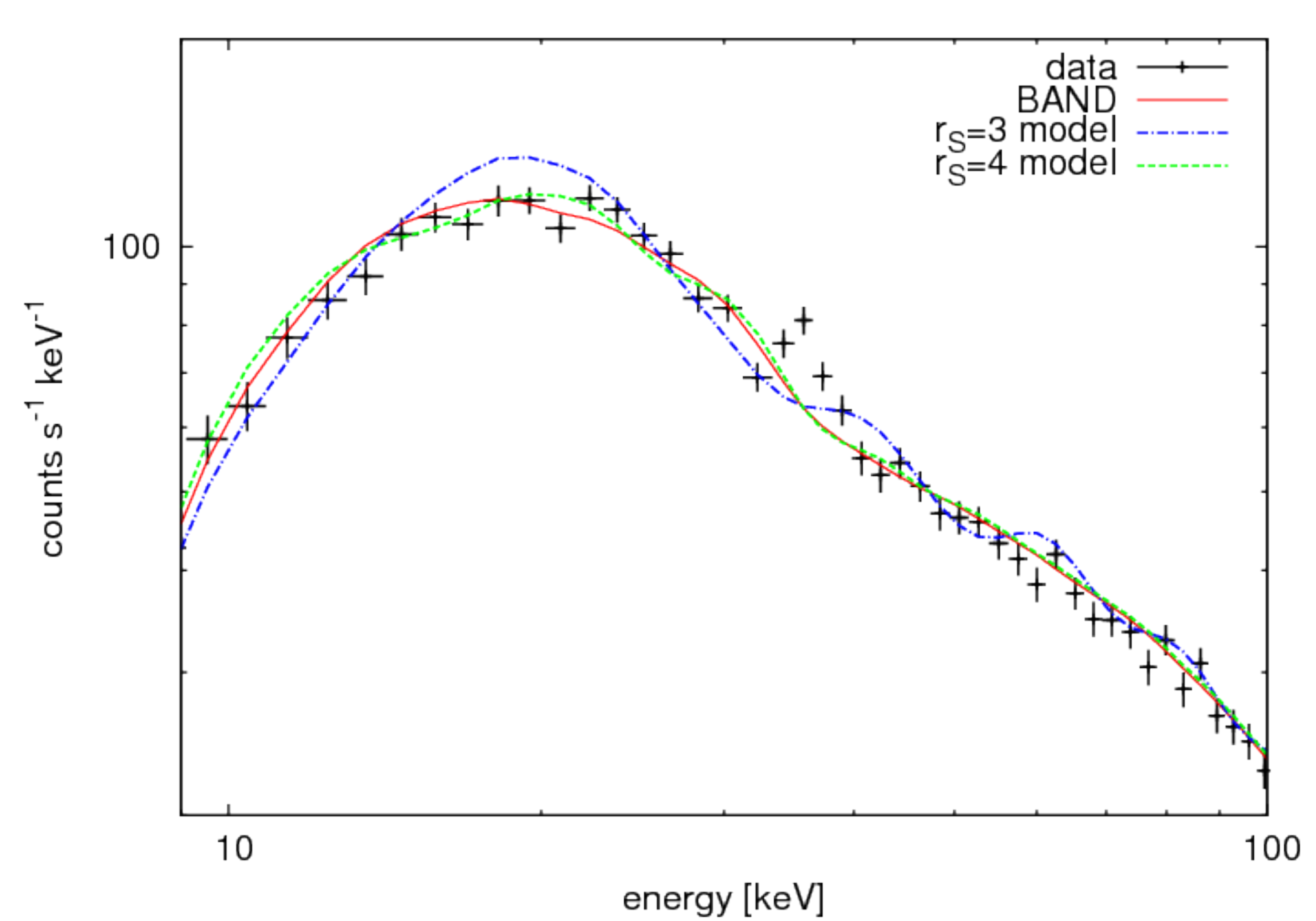}%
   \caption{\label{rS34} The spectrum of GRB 090424592 using NaI
     detector n7. The BAND and femtolensing fits are superimposed. 
     The parameters are $r_{S} = 3$, $4$, and lens mass $5\times 10^{18} \,$g.
     The excess at $33\,$keV (K-edge) is an instrumental effect
     seen on many bright bursts. }
 \end{figure}

The data are then modified by incorporating the spectral 
fringe patterns for a range of lens masses $M$ and source positions $r_{S}.$  
The simulated data and the corresponding femtolensing fit are presented in Fig.~\ref{MD_femto}.
Neither  BKN nor BAND models are able to fit the simulated data (see Fig.~\ref{MD_BAND}). 
The values of $r_{S}$ are then changed until the $\chi^2$ of the fit obtained is not
significantly different from the $\chi^2$ of the unmodified data.  
More precisely, the $\chi^2$ difference $\Delta \chi^2$ should be distributed
in the large sample limit as a $\chi^2$ distribution with 2 degrees of freedom
according to  Wilk's theorem \cite{1996ApJ...461..396M}. The value $\Delta \chi^2 = 5.99,$ which corresponds to a $\chi^2$ probability of 5\% for 2 d.o.f, was taken as the cut value. 
The effect of changing $r_{S}$  on the femtolensing model is illustrated on Fig.~\ref{rS12} and \ref{rS34}.

In Fig.~\ref{effR} we show the maximum and minimum detectable $r_S$ for different lens masses. 
The maximum difference between $r_{S,max}$ and $r_{S,min}$ appears at $M = 5  \times 10^{18}\,$g,
which indicates the maximum of femtolensing cross-section.

\section{Results\label{sec:results}}
The 20 burst sample from Tab.~\ref{table} have been fitted with the standard BKN, BAND and SBKN models. 
The models with the best $\chi^2$ probability were selected and are shown on Tab.~\ref{table}. 
The bursts are well fitted by these standard models, so that there is no evidence for femtolensing in the data. 

\begin{figure}
 \includegraphics[width=6cm,angle=-90]{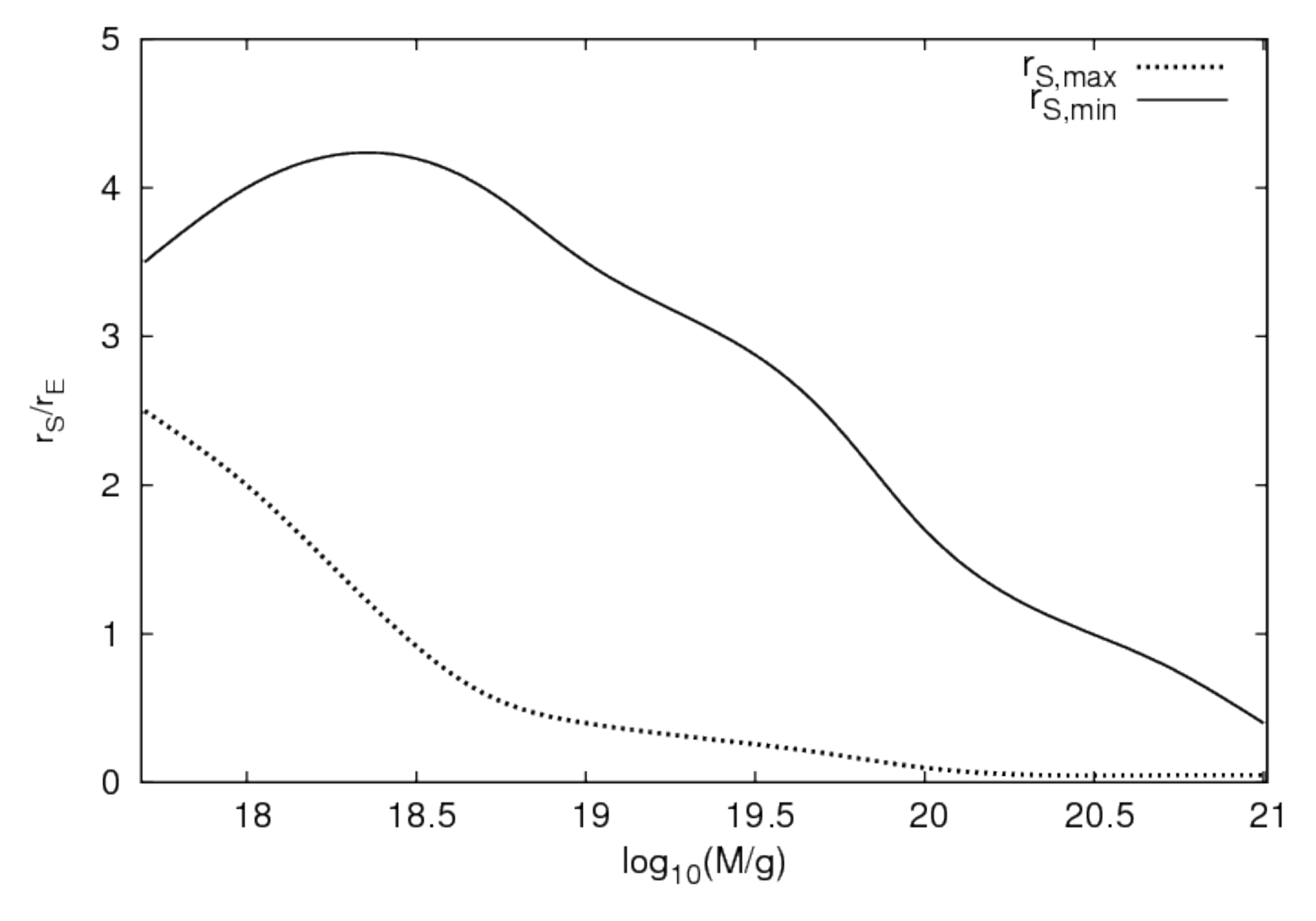}%
 \caption{\label{effR} Minimum and maximum detectable $r_S/r_E$ as a function of lens mass for GRB 090424592.  }
 \end{figure}

As explained in section \ref{sec:probability}, the lensing probability 
for each burst depends on the lens mass and on the $r_{S,min}$ and $r_{S,max}$ values. Since the sensitivity
of GBM to femtolensing is maximal for lens masses of $\sim 5  \times 10^{18}\, $ g (see Fig.~\ref{effR}), 
the values of $ r_{S,min}$ and $r_{S,max}$ for each event were first determined at a mass $M = 5  \times 10^{18}\, $ g by simulation.  
As explained in Sec.~\ref{sec:probability}, the value of $r_{S,min}$ is set by the period of the spectral fringes 
so that it is independent of the burst luminosity. 
The values of  $r_{S,max}$ obtained are listed  in Tab.~\ref{table}.
The lensing probability is  then calculated for both  the FRLW and Dyer \& Roeder 
cosmological models using each burst redshift, the most probable lens position 
and the values of $r_{S,min}$ and $r_{S,max}$ for the mass $M = 5  \times 10^{18}\,$g. 
The number of expected lensed bursts in the sample is the sum of the lensing probabilities. 
It depends linearly on $\Omega_{CO}.$

\begin{figure}
 \includegraphics[width=6cm,angle=-90]{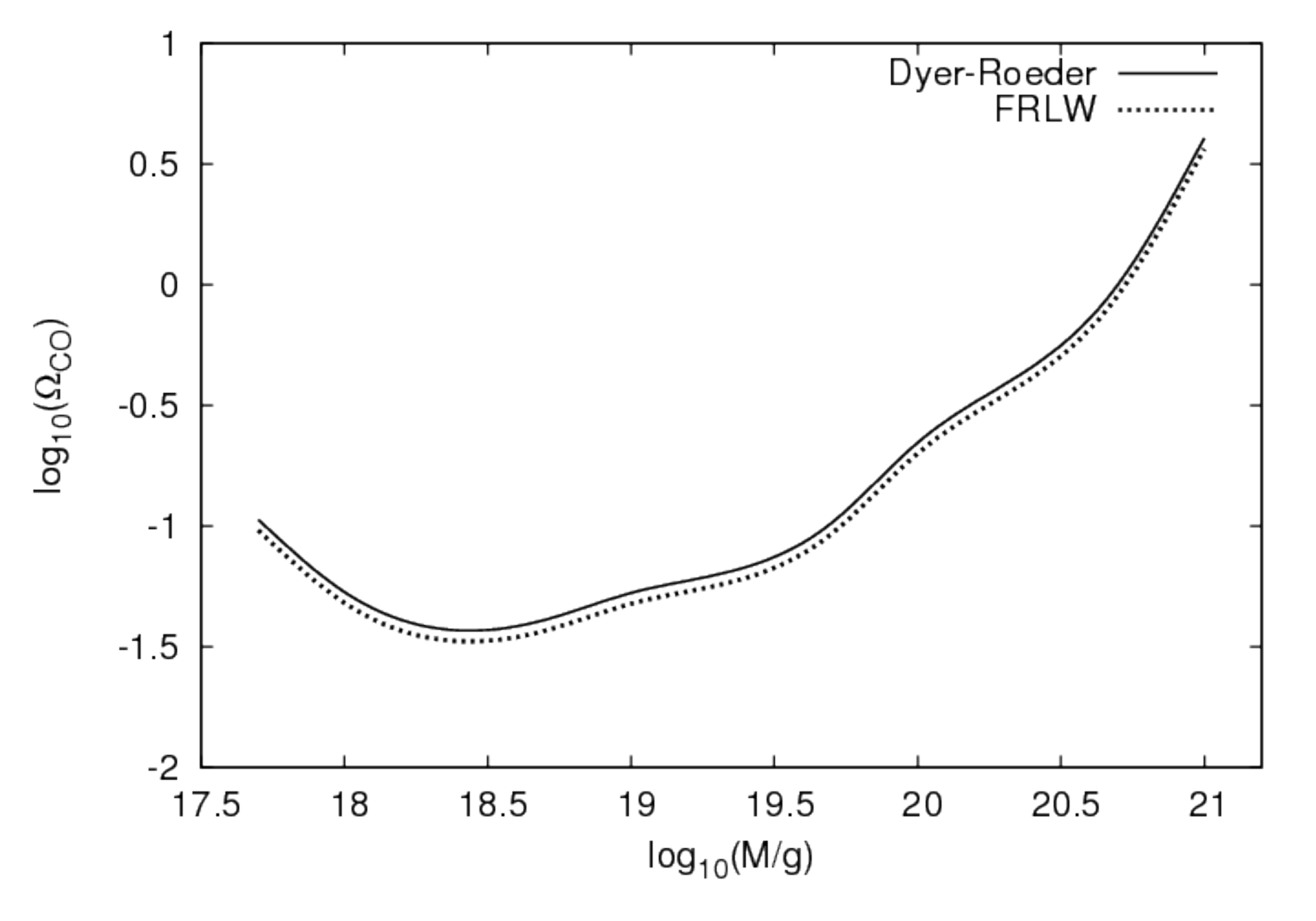}%
 \caption{\label{fraction} Constraints on the fraction (or normalize density) of compact objects. 
 The zones above the curves are excluded at the 95\% confidence level.}
 \end{figure}

Since no femtolensing is observed, the number of expected events 
should be less than 3 at 95\% confidence level (C.L.).
The constraints on  the density of compact objects $\Omega_{CO} $  
is derived to be less than 4\% at 95\% C.L for both cosmological models.
The values of the lensing probabilities for all the bursts in our sample 
assuming the constrained density of compact objects are shown in Tab.~\ref{table}. 
The limits at other lens masses are obtained by normalizing the $\Omega_{CO} $ at $M = 5\times 10^{18} \mbox{g}$ by the cross section $\sigma$. 
The cross section  is calculated 
using the Eq.~\ref{sigma} and the values of $r_{S,min}$ and $r_{S,max}$  from Fig.~\ref{effR}.  
The limits on $\Omega_{CO}$   at $95\%\,$C.L. are plotted in Fig.~\ref{fraction}.


   
\section{Discussion and conclusions\label{sec:conclusions}}
Cosmological constraints on the PBH abundance are reviewed by Carr et al.~\cite{2010PhRvD..81j4019C}. 
One way to obtain the abundance of PBH is to constrain the density of compact objects $\Omega_{CO}$.  
Note that the limits on the compact object abundance  in the $10^{26} - 10^{34}\,$g range obtained with microlensing are at the 1$\%$~level.

It is stated in Abramowicz et al.~\cite{2009ApJ...705..659A}
that the mass range $10^{16}\,$g $< M_{BH} < 10^{26}\,$g is virtually unconstrained. 
Under this range, the $\Omega_{CO}$ is constrained by PBH evaporation. 
Above this range, the constraints come from microlensing. 
The new idea by Griest et al.~\cite{2011PhRvL.107w1101G} shows that the microlensing limit could be improved and
get constraints down to $10^{20}\,$g with the Kepler satellite observations.Ó

The FERMI satellite  was launched three and a half years ago.
Since then, almost 1000 of GRB were observed with the GBM detector. 
In many cases data quality is good enough to reconstruct  time-resolved spectra. 
This unique feature is exploited in our femtolensing search by selecting the first few seconds of a burst in data analysis.

Our limits were obtained by selecting only those bursts with known redshifts in the GBM data. 
This reduces the data sample from the 500 bursts detected in the first 2 years to only 20.
The constraints on $\Omega_{CO}$   obtained at  the $95\%\,$ C.L. are shown on Fig.~\ref{fraction}.
These constraints improve the existing constraints by a factor of 4 in the mass range $5 \times 10^{17} $ -- $10^{20}\,$g.

After ten years of operation, the GBM detector should collect over 2500 bursts. 
Only a few of the bursts, say 100, will have a measured redshift and sufficient spectral coverage.
By applying the methods described in this paper, our limits will then improve by a factor of 5 reaching 
a sensitivity to density of compact objects down to the 1\% level.  


\begin{table*}[H] 
 \caption{\label{sample} The sample of 20 GBM GRBs used in the analysis.} 
 \begin{ruledtabular}
 \begin{center}
 \begin{tabular}{cccllccccc}
          &                          &               &   
                                                                           &  \multicolumn{2}{c}{Fit to simulated data}  
                                                                                                                        &                 &       &                & \\ 
 Name          
                                       & $z_S$ &   \multicolumn{2}{c}{Fit to the data \footnote{Fit has been performed using only the photons arrived in less than 10s from the beginning of the burst.}}             
                                                                          & Model   &Femtolensing  
                                                                                                                         & $r_{S,max}$&$z_L$&\multicolumn{2}{c}{ Lensing Probability }  \\
                                     &                 & Model     & $\chi^2$/d.o.f  
                                                                                             &  $\chi^2$/d.o.f &$\chi^2$/d.o.f
                                                                                                                                   	& 
	                                                                                                                                           &     & FRLW \footnote{for assumed $\Omega_{CO}=0.037$}  & Dyer-Roeder \footnote{for assumed $\Omega_{CO}=0.041$}\\
 \hline
GBM 080804972	&2.2045   &BAND	&68/74	&116/63		&65/61	&2   &0.770	& 0.110 & 0.112 \\
GBM 080916009C   &4.3500	&BKN	&68/75	&91/86		&78/84	&3   &1.087	& 0.580 & 0.539 \\
GBM 080916406A   &0.6890	&BKN	&58/57	&122/87		&110/85	&3   &0.324	& 0.036 & 0.040 \\
GBM 081121858	&2.5120	&BKN	&39/49	&52/49		& 42/47	&3   &0.829	& 0.296 & 0.298 \\
GRB 081222204 	&2.7000   &BKN	&73/66	&82/62		&63/60      &3   &0.859	& 0.326 & 0.325 \\ 
GRB 090102122	&1.5470	&BAND	&81/85	&103/85		&93/83	&3   &0.603	& 0.146 & 0.154 \\
GRB 090323002	&3.5700	&BAND	&77/77	&121/64		&90/62	&2   &0.964	& 0.206 & 0.197 \\
GRB 090328401	&0.7360	&BKN	&105/70	&123/70		&103/68	&2   &0.346	& 0.018 & 0.020 \\
GRB 090424592	&0.5440     &BAND	&78/67	&115/75		&97/73	&4   &0.256	& 0.042 & 0.046 \\
GRB 090510016	&0.9030	&BKN	&62/66	&173/98		&158/96	&1.5&0.406	& 0.015 & 0.016 \\
GRB 090618353	&0.5400	&BAND	&59/72	&79.5/72		&66/70	&3    &0.254	& 0.023 & 0.025 \\
GRB 090926181	&2.1062	&BAND	&87/81	&105/81		&93/79	&4    &0.737	& 0.413 & 0.423 \\
GRB 091003191	&0.8969	&BKN	&93/94	&105/94		&94/92	&3    &0.400  	& 0.059 & 0.064 \\
GRB 091020900	&1.7100	&BKN	&74/69	&116/69		&90/67	&2.5 &0.667	& 0.119 & 0.124 \\
GRB 091127976	&0.4900	&BAND	&78/74	&84/74		&76/72	&4    &0.240	& 0.034 & 0.037 \\
GRB 091208410	&1.0630	&BAND	&55/55	&115/59		&56/57	&2.5 &0.457	& 0.055 & 0.059 \\
GRB 100414097	&1.3680	&BKN	&65/61	&97.5/68		&86/66	&2.5 &0.560	& 0.084 & 0.089 \\
GRB 100814160A	&1.4400	&BKN	&42/40	&138/42		&128/40 	&2    &0.590	&  0.058 & 0.061 \\
GRB 100816009	&0.8049	&BKN	&73/52	&95/50		&66/48	&2.5 &0.360	&  0.034 & 0.037 \\
GRB 110731465	&2.8300	&SBKN	&72/64	&125/64		&97/62	&3    &0.877	&  0.347 & 0.344 \\
 \end{tabular}
 \end{center}
 \end{ruledtabular}
\label{table}
\end{table*}


\begin{acknowledgments}
Part of this research was supported by the Polish Ministry of Science
and Higher Education under grant no. DEC-2011/01/N/ST9/06007 and grant no. ERA-NET-ASPERA/01/10.
\end{acknowledgments}

\bibliography{femto}
\end{document}